\documentclass[twocolumn,showpacs,preprintnumbers,showkeys,superscriptaddress,prc]{revtex4}
\usepackage{CJK}
\usepackage{mathrsfs}
\usepackage{amssymb}
\usepackage{amsmath}
\usepackage{graphicx}
\usepackage{dcolumn}
\usepackage{bm}
\usepackage{graphicx}
\usepackage{float}
\usepackage[normalem]{ulem}
\usepackage{color}
\usepackage{multirow}

\newcolumntype{d}[1]{Dc{.}{.}{#1}}
\begin{document}
\begin{CJK*}{UTF8}{}

\title{Global correlations between electromagnetic and spectroscopic properties of  collective $2^+_1$ and $2^+_2$ states}
\author{Z. Z. Qin ({\CJKfamily{gbsn}覃珍珍})}
\email{qin_zhenzhen@hotmail.com}
\affiliation{School of Science, Southwest University of Science and Technology, Mianyang 621900, China}
\author{Y. Lei ({\CJKfamily{gbsn}雷杨})\footnote{corresponding author: leiyang19850228@gmail.com}}
\affiliation{Key Laboratory of Neutron Physics, Institute of Nuclear Physics and Chemistry, China Academy of Engineering Physics, Mianyang 621900, China}
\author{S. Pittel}
\email{pittel@bartol.udel.edu}
\affiliation{Bartol Research Institute and Department of Physics and Astronomy, University of Delaware, Newark, Delaware 19716, USA}
\date{\today}

\begin{abstract}
By using the general triaxial rotor model (TRM) and the phonon-configuration mixing scheme within an anharmonic-vibrator(AHV) framework, a series of global correlations between electromagnetic properties of nuclear $2^+_1$ and $2^+_2$ states are analytically established. The correlations from both models can roughly describe the experimental data involving quadrupole collectivity with few exceptions. Furthermore, there seems to be a robust orthogonal transformation between the AHV and TRM bases for realistic nuclear systems, suggesting that the two models may in fact be describing the collective features of nuclear low-lying states in similar model spaces.
\end{abstract}
\maketitle
\end{CJK*}

\section{Introduction}
The quadrupole collectivity exhibited by low-lying states in atomic nuclei robustly maintains rotational characteristics. This includes a large range of nuclei for which the low-lying levels do not behave like those of an axially-symmetric rigid rotor, as they have energy ratios of $R=E(4^+_1)/E(2^+_1)$ that deviate, often substantially, from 10/3. For example, most doubly-even nuclei with $A>56$ for which $R=E(4^+_1)/E(2^+_1)<3$ and B(E2, $2^+_1\rightarrow 0^+_1$)$<100$ W.U. have ratios of quadrupole transition rates B(E2, $4^+_1\rightarrow 2^+_1$)/B(E2, $2^+_1\rightarrow 0^+_1$) near 10/7, which is the axially-symmetric rigid rotor limit \cite{rowe}. Furthermore, theoretical shell-model calculations, with both effective interactions and random interactions, tend to provide the rotational Alaga ratio of $Q^2(2^+_1)/{\rm B(E2,~}2^+_1\rightarrow 0^+_1)=64\pi/49$, regardless of the low-lying spectrum behavior \cite{alaga-1,alaga-2,alaga-3}.

A recent experimental survey \cite{allmond} of quadrupole moments of the lowest two $I^{\pi}=2^+$ states [denoted by $Q(2^+_1)$ and $Q(2^+_2)$, respectively] demonstrated a global $Q(2^+_1)= -Q(2^+_2)$ correlation, {\it i.e.} another rotor-like correlation proposed therein, across a wide range of masses and deformations accompanied by $R$ values between 2 (the vibrational limit) and 10/3.  One of the present authors also observed a robust $Q(2^+_1)=-Q(2^+_2)$ correlation in shell-model random-interaction ensembles without rotational yrast states \cite{lei}.

It is important to note here that only the pure axially symmetric rotor requires $R=10/3$.  Several more general nuclear rotor models can produce ``realistic" low-lying spectra away from this limiting behavior. This includes models with a strong microscopic underpinning, such as the  coupled-SU(3) model \cite{rowe} and  the Sp(3,R) model \cite{sp3r}, as well as others that are more phenomenological, such as the triaxial rotor model (TRM) \cite{trm}.  One of these models, the TRM with three E2-tensor-independent inertia for the three different principal axes \cite{trm-first}, trivially gives rise to the $Q(2^+_1)=-Q(2^+_2)$ correlation without limits on the $R$ value. This approach, which adopts five model parameters and an analytical formalism, has been used extensively for the description of experimental data on E2 collectivity \cite{trm-os,trm-pt,trm-er,trm-ge}.

It is also possible that the $Q(2^+_1)= -Q(2^+_2)$ correlation might be present in non-rotor models. Here we will consider the possibility of describing it through phonon-configuration mixing in the anharmonic-vibrator model (AHV) \cite{ahv-1,ahv-2,ahv-3}. In the early 90s, Casten {\it et al.} discussed  the linear correlation between $E(4^+_1)$ and $E(2^+_1)$  of nuclei in the AHV model, showing that the model could be applied to nuclei with $R=2.05\sim 3.15$ \cite{casten-1}. Thus, the AHV model likewise has a much weaker constraint on the $R$ value than the axially symmetric rotor model, while still describing quadrupole collective features. Therefore, it too might provide a spectrally-consistent explanation for the global $Q(2^+_1)= -Q(2^+_2)$ correlation, while perhaps also providing other correlations between electromagnetic properties and spectra. However, Ref. \cite{casten-1} only focused on the AHV behavior of the yrast band, and thus did not consider the effects of phonon-configuration mixing. The experimental evidence for the global existence of phonon-configuration mixing in $2^+_1$ and $2^+_2$ states is not well established yet, to the best of our knowledge, but is nevertheless worth exploring.

This work aims to examine the ability of the TRM and phonon-configuration mixing in the AHV to describe global correlations between nuclear low-lying electromagnetic properties and level properties. First, in Sec. \ref{fra}, we review the formalism of the TRM and AHV, which will be used to derive possible correlations among excitation energies, E2 transition rates and electromagnetic moments. We then report in Sec. \ref{sys} an experimental survey based on the ENSDF database \cite{ensdf} to verify the applicability of the correlations derived from the previous step for the two models. Finally, in  Sec. \ref{cor}, we discuss the possibility that there is in fact an underlying relation between these two seemingly different views of nuclear collectivity, the TRM and the AHV. In Sec. \ref{sum} we summarize the main features and conclusions of our study.

\section{Model frameworks}\label{fra}
\subsection{The TRM}
Details on the formalism of the TRM with independent inertia and electric quadrupole tensors were presented in Ref. \cite{trm-first}. Here, we only present a few key formulas that are needed for this work. In the TRM, the Hamiltonian matrixes for $2^+$ and $4^+$ states, respectively, are written schematically as:
\begin{equation}\label{trm-h}
\begin{aligned}
H^{2^+}_{\rm TRM}&=
\begin{pmatrix}
6A&4\sqrt{3}G\\
4\sqrt{3}G&6A+4F\\
\end{pmatrix}\\
H^{4^+}_{\rm TRM}&=
\begin{pmatrix}
20A&12\sqrt{5}G&0\\
12\sqrt{5}G&20A+4F&4\sqrt{7}G\\
0&4\sqrt{7}G&20A+16F\\
\end{pmatrix},\\
\end{aligned}
\end{equation}
where $A$, $G$ and $F$ are Hamiltonian parameters related to the three components of the inertia tensor.

The $2^+$ states are orthogonal combinations of the $K=0$ and $K=2$ configurations, according to
\begin{equation}\label{trm-2+}
\begin{aligned}
|2^+_1\rangle&=\cos\Gamma |K=0\rangle -\sin\Gamma|K=2\rangle\\
|2^+_2\rangle&=\sin\Gamma |K=0\rangle+\cos\Gamma|K=2\rangle ~,
\end{aligned}
\end{equation}
where $K$ is the projection of the angular momentum with respect to the intrinsic coordinate system, and $\tan2\Gamma=2\sqrt{3}G/F$ defines the $K$ mixing.

The electric quadrupole properties of the lowest $2^+$ states can be  expressed as
\begin{equation}\label{trm-q}
\begin{aligned}
{\rm B(E2,}2^+_1\rightarrow 0^+_1)&=\frac{Q^2_0}{16\pi}\cos^2(\gamma+\Gamma)\\
{\rm B(E2,}2^+_2\rightarrow 0^+_1)&=\frac{Q^2_0}{16\pi}\sin^2(\gamma+\Gamma)\\
{\rm B(E2,}2^+_2\rightarrow 2^+_1)&=\frac{5Q^2_0}{56\pi}\sin^2(\gamma-2\Gamma)\\
Q(2^+_1)=-\frac{2}{7}~Q_0&\cos(\gamma-2\Gamma)=-Q(2^+_2) ~,
\end{aligned}
\end{equation}
where $Q_0$ is the static quadrupole moment, and $\gamma$ is a parameter related to the nuclear quadrupole deformation. One sees that a $Q(2^+_1)=-Q(2^+_2)$ correlation is unconditionally conserved in the TRM.

\subsection{The AHV}
In the AHV description \cite{ahv-2}, the model space of $2^+_1$ and $2^+_2$ states can be constructed from one-phonon and two-phonon excitations of the phonon vacuum $|0 \rangle$ (namely the $0^+$ ``ground state") denoted by  $|1\rangle$ and $|2\rangle$, respectively. In this model space, the Hamiltonian matrix is written as
\begin{equation}\label{ahv-h}
H^{2^+}_{\rm AHV}=
\begin{pmatrix}
\hbar\omega & \lambda\\
\lambda & 2\hbar\omega
\end{pmatrix},\\
\end{equation}
where $\hbar\omega$ and $\lambda$ are the one-phonon excitation energy and the mixing energy between the phonon configurations, respectively. The $2^+_1$ and $2^+_2$ states of an AHV nucleus correspond to a mixing of the two configurations, defined by
\begin{equation}\label{ahv-2+}
\begin{aligned}
|2^+_1\rangle&=a_1 |1\rangle+a_2|2 \rangle\\
|2^+_2\rangle&=-a_2 |1\rangle+a_1 |2 \rangle ~, \\
\end{aligned}
\end{equation}
where $a_1$ and $a_2$ are amplitudes derived from diagonalization of Eq. (\ref{ahv-h}) with the normalization constraint, $a_1^2+a_2^2=1$. The electric quadrupole operator in the AHV \cite{ahv-q} is given by
\begin{equation}
\hat{Q}=\chi(b^{\dagger}+\tilde{b}),
\end{equation}
where $b^{\dagger}$ and $\tilde{b}$ are creation and (time-reversed) annihilation operators of the phonon, respectively, and $\chi$ is a free parameter. The quadrupole properties in this model are given by
\begin{equation}\label{ahv-q}
\begin{aligned}
{\rm B(E2,}2^+_1\rightarrow 0^+_1)&=\frac{\chi^2a^2_1}{5}\langle 0||\tilde{b}||1\rangle^2\\
{\rm B(E2,}2^+_2\rightarrow 0^+_1)&=\frac{\chi^2a^2_2}{5}\langle 0||\tilde{b}||1\rangle^2\\
{\rm B(E2,}2^+_2\rightarrow 2^+_1)&=\frac{\chi^2(a^2_1-a^2_2)^2}{5}\langle 1||\tilde{b}||2\rangle^2\\
Q(2^+_1)=\frac{8\chi a_1a_2}{5}&\sqrt{\frac{2\pi}{7}}\langle 1||\tilde{b}||2\rangle=-Q(2^+_2)~.
\end{aligned}
\end{equation}
Again, we have a $Q(2^+_1)=-Q(2^+_2)$ correlation regardless of the spectral behavior.

We should emphasize that the $Q(2^+_1)=-Q(2^+_2)$ correlation only emerges when the $2^+$ model space is constructed solely in terms of the $|1\rangle$ and $|2\rangle$ states. In principle, there may be multi-phonon mixing in the $2^+_1$ and $2^+_2$ states, which can distort this correlation. Since the $Q(2^+_1)=-Q(2^+_2)$ correlation seems to be present in experiment \cite{allmond}, we conclude that these multi-phonon configurations are probably not very important in realistic nuclei. In what follows, we will therefore neglect them in our AHV analysis, and assume an appropriately truncated version of the AHV model.

\section{Systematic comparison with experimental data}\label{sys}
\subsection{Spectral properties}\label{spe}
Diagonalization of Eq. (\ref{trm-h}) for given values of $A$, $F$ and $G$ provides excitation energies of $2^+_1$, $2^+_2$ and $4^+_1$ states in the TRM approach. This diagonalization is equivalent to solving quadratic and cubic equations. Conversely, one can take $A$, $G$ and $F$ in Eq. (\ref{trm-h}) as unknown variables, and experimental excitation energies as input parameters. Roots of such equation would then give appropriate $A$, $G$ and $F$ parameters for each specific nucleus. We do not present details on the mathematical process followed here. However, we note that there may be no real-number roots of $A$, $G$ and $F$ for some nuclei, in which case, Eq. (\ref{trm-q}) can not provide observable E2 transition rates and moments. This could happen if the experimental low-lying level scheme is incomplete, or if the TRM is inappropriate for the nucleus under investigation.

\begin{figure}
\includegraphics[angle=0,width=0.45\textwidth]{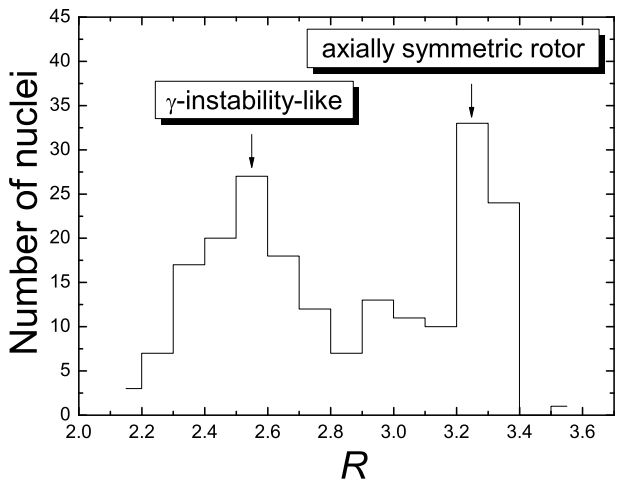}
\caption{ $R=E(4^+_1)/E(2^+_1)$ distribution of the 203 TRM-solvable nuclei in the ENSDF (see text). Peaks for $\gamma$-instability  ($R\simeq 5/2$) and an axially-symmetric rotor ($R\simeq10/3$) are highlighted.}\label{trm-r}
\end{figure}

In the ENSDF, there are 203 TRM-solvable nuclei.  In Fig. \ref{trm-r}, we present the $R=E(4^+_1)/E(2^+_1)$ distribution of these  nuclei. One sees that this distribution spreads over the whole $R>2$ region, indicating that the TRM could indeed provide non-rotor-like spectra, but still maintain the rotational $Q(2^+_1)=-Q(2^+_2)$ correlation demonstrated by Eq. (\ref{trm-q}). We also note that there are two peaks in this distribution around $R=5/2$ and $R=10/3$. The $R=10/3$ peak obviously corresponds to rotational nuclei with axially symmetric deformation. The $R=5/2$ peak, on the other hand, is normally taken as the sign of the O(6) limit in the interacting boson model (IBM) \cite{ibm}, corresponding to $\gamma$-instability. It has been pointed out that the TRM may share a similar structural pattern for $K=0$ and $K=2$ bands with a $\gamma$-unstable model \cite{trm-os}, in agreement with the $R=5/2$ predominance exhibited here.

\begin{figure}
\includegraphics[angle=0,width=0.45\textwidth]{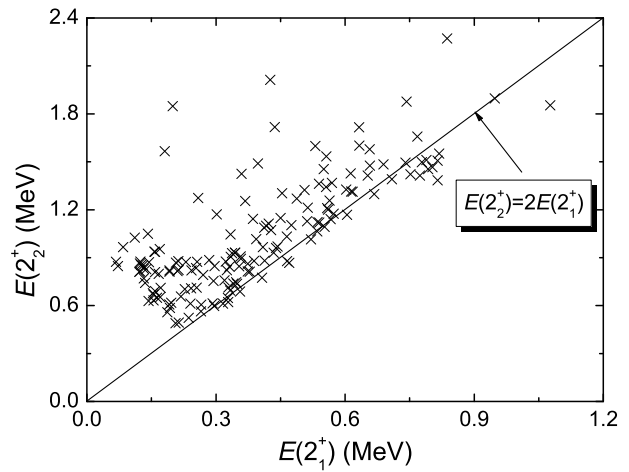}
\caption{Experimental $E(2^+_2)$ versus $E(2^+_1)$ for the  177 nuclei with $R=2.05\sim 3.15$ in the ENSDF. The $E(2^+_2)=2E(2^+_1)$ line is highlighted.}\label{ahv-e}
\end{figure}

The spectral structure of nuclei using the AHV approach is simpler than that using the TRM.  As demonstrated by Eq. (\ref{ahv-2+}), $2^+_1$ and $2^+_2$ states are the mixing of one-phonon and two-phonon configurations. According to perturbation theory, $E(2^+_1)$ is below the one-phonon excitation energy, {\it i.e.} $\hbar\omega$, while $E(2^+_2)$ is  beyond the two-phonon excitation energy, {\it i.e.} $2\hbar\omega$. In Fig. \ref{ahv-e}, we plot all the available experimental data on nuclei with $E(2^+_2)$ versus $E(2^+_1)$ in the range $R=2.05\sim 3.15$, which were already assigned to the AHV by Casten {\it et al.}. This experimental ensemble includes 177 nuclei in the ENSDF. One sees that most $R=2.05\sim 3.15$ nuclei fall beyond the $E(2^+_2)=2E(2^+_1)$ line, which forms a sharp edge in Fig. \ref{ahv-e}. This observation suggests th
at the ensemble of $R=2.05\sim 3.15$ nuclei is indeed a reasonable sample of AHV nuclei, with phonon-configuration mixing robustly existing in the $2^+_1$ and $2^+_2$ states, in agreement with the classification of Casten {\it et al.}.

\subsection{Correlation between $\mu(2^+_1)$ and $\mu(2^+_2)$}\label{subsec-mu}
To calculate magnetic moments ($\mu$) within the TRM framework, we adopt the following schematic $\mu$ matrix elements for a rigid rotor with good quantum numbers $IK$ \cite{bm}:
\begin{equation}\label{trm-mu}
\begin{aligned}
\langle IK=0|\hat{\mu}|IK=0\rangle&=g_RI\\
\langle IK=2|\hat{\mu}|IK=2\rangle&=g_RI+(g_K-g_R)\frac{K^2}{I+1} ~.\\
\end{aligned}
\end{equation}
Here, $\hat{\mu}$ is the magnetic-moment operator;  $g_R$ and $g_K$ are the $g$ factors for rotational motion and intrinsic motion, respectively. The exchange matrix $\langle IK=0|\hat{\mu}|IK=2\rangle$ vanishes, since $\hat{\mu}$ is a rank-one vector. For a uniformly charged liquid drop, protons and neutrons have the same contribution to the total nuclear angular momentum in the laboratory frame, but only the protons contribute to the magnetic moment.  Thus, on average, $g_R\simeq Z/A$ \cite{bm}, where $Z$ is the proton numbe
r and $A$ is the mass number. The term $g_K$ in Eq. (\ref{trm-mu}) represents the magnetic moment from quasi-particle excitations in the intrinsic frame. In the asymptotic limit, $g_K$ equals 1 for the quasi-proton configuration, and 0 for the quasi-neutron configuration \cite{gk}. In TRM-applicable nuclei with significant deformation, valence proton and neutron configurations are
strongly mixed \cite{np}, which implies that each valence proton or neutron has the same probability of being excited. Thus, $g_K$ can be estimated to be $g_K=N_p/(N_p+N_n)$,
where $N_p$ and $N_n$ are the valence proton and neutron numbers, respectively, from which the magnetic moments of $2^+$ states defined by Eq. (\ref{trm-2+}) within the TRM  are given by
\begin{equation}
\begin{aligned}
\mu(2^+_1)&=2{\frac{Z}{A}\sin^2\Gamma + [\frac{2Z}{3A}+\frac{4N_p}{3(N_p+N_n)}]}\cos^2\Gamma\\
\mu(2^+_2)&=2{\frac{Z}{A}\cos^2\Gamma +[\frac{2Z}{3A}+\frac{4N_p}{3(N_p+N_n)}]}\sin^2\Gamma ~.
\end{aligned}
\label{trm-mu-az}
\end{equation}

In the AHV, the first-order approximation to the magnetic dipole moment operator\cite{ahv-q} in a sharp-edged liquid drop model may be simplified as
\begin{equation}
\hat{\mu}=\eta[(b^{\dagger}+\tilde{b})\times (b^{\dagger}+\tilde{b})]^1~,
\end{equation}
where $\eta$ is a constant for a specific nucleus determined b
y the nuclear charge distribution; the superscript $1$ is the angular momentum of this operator, {\it i.e.} the two $b^{\dagger}+\tilde{b}$ operators are coupled to a rank-one dipole operator. The calculation of the magnetic moments of $2^+$ states defined by Eq. (\ref{ahv-2+}) can be simplified as follows. First, the irreducible matrix elements $\langle 1||[b^{\dagger}b^{\dagger}]^1||1\rangle$,  $\langle 2||[b^{\dagger}b^{\dagger}]^1||2\rangle$, $\langle 1||[\tilde{b}\tilde{b}]^1||1\rangle$ and $\langle 2||[\tilde{b}\tilde{b}]^1||2\rangle$ vanish due to phonon-number conservation.  Second, the coupled boson commutation relation requires $[b^{\dagger}\tilde{b}]^1=[\tilde{b}b^{\dagger}]^1$ \cite{com}. Third, $[\tilde{b}b^{\dagger}]^1=\hat{L}/\sqrt{10}$, where $\hat{L}$ is the boson angular momentum operator. Finally, since the phonon configurations $|1\rangle$ and $|2\rangle$ both have boson angular momentum $L=2$ as a good quantum number, $\mu(2^+_1)$ and $\mu(2^+_2)$ in the AHV model are given by
\begin{equation}
\begin{aligned}
\mu(2^+_1)&=\frac{2}{\sqrt{10}}~\eta(a_1^2+a_2^2)=\frac{2}{\sqrt{10}}~\eta\\
&=\mu(2^+_2) ~.
\end{aligned}
\end{equation}
Thus, the magnetic moments are correlated in the AHV  as $\mu(2^+_1)=\mu(2^+_2)$.

\begin{table}
\caption{Experimentally available $\mu(2^+_1)$ and $\mu(2^+_2)$ val
ues \cite{ensdf} compared with our TRM estimates. The ratios of $\mu(2^+_2)/\mu(2^+_1)$ are listed to demonstrate the $\mu(2^+_1)=\mu(2^+_2)$ correlation.  Nuclei with no TRM estimates here are insolvable in the TRM framework, as discussed in Sec. \ref{spe}.}\label{mu}
\begin{tabular}{lllllllllllllllllllllllllllllllll}
\hline\hline
\multirow{2}{*}{Nucleus}	&		&	\multicolumn{3}{c}{Expt.}					&		&	\multicolumn{2}{c}{TRM}			\\
\cline{3-5}\cline{7-8}															
	&	$~~~$	&	$\mu(2^+_1)$	&	$\mu(2^+_2)$	&	$\frac{\mu(2^+_2)}{\mu(2^+_1)}$	&	$~~~$	&	$\mu(2^+_1)$	&	$\mu(2^+_2)$	\\
\hline															
$^{80}$Kr	&		&	+0.76(10)	&	+1.3(7)	&	1.7(9)	&		&		&		\\
$^{86}$Sr	&		&	+0.57(3)	&	+0.8(3)	&	1.4(5)	&		&		&		\\
$^{92}$Zr	&		&	-0.360(20)	&	+1.5(10)	&	-4(3)	&		&		&		\\
$^{132}$Xe	&		&	+0.651(24)	&	+0.2(4)	&	0.3(6)	&		&		&		\\
$^{150}$Sm	&		&	+0.77(5)	&	+0.72(17)	&	0.9(2)	&		&	0.81	&	0.62	\\
$^{152}$Sm	&		&	+0.82(4)	&	+0.76(19)	&	0.9(2)	&		&	0.82	&	0.58	\\
$^{160}$Dy	&		&	+0.723(19)	&	+0.80(5)	&	1.11(8)	&		&	0.83	&	0.63	\\
$^{162}$Dy	&		&	+0.686(28)	&	0.92(6)	&	1.3(1)	&		&	0.82	&	0.62	\\
$^{164}$Dy	&		&	+0.684(23)	&	0.76(6)	&	1.1(1)	&		&	0.81	&	0.60	\\
$^{166}$Er	&		&	+0.641(10)	&	0.69(8)	&	1.1(1)	&		&	0.82	&	0.64	\\
$^{168}$Er	&		&	+0.
642(12)	&	+0.72(14)	&	1.1(2)	&		&	0.81	&	0.62	\\
$^{184}$W	&		&	+0.578(14)	&	+0.25(8)	&	0.4(1)	&		&	0.80	&	0.65	\\
$^{186}$W	&		&	+0.615(24)	&	+0.39(8)	&	0.6(1)	&		&	0.80	&	0.64	\\
$^{188}$Os	&		&	+0.596(22)	&	+0.78(7)	&	1.3(1)	&		&	0.81	&	0.67	\\
$^{190}$Os	&		&	+0.692(30)	&	+0.66(8)	&	1.0(1)	&		&	0.80	&	0.66	\\
$^{192}$Os	&		&	+0.792(20)	&	+0.58(4)	&	0.73(5)	&		&	0.78	&	0.65	\\
$^{192}$Pt	&		&	+0.590(18)	&	+0.61(8)	&	1.0(1)	&		&		&		\\
$^{194}$Pt	&		&	+0.60(3)	&	+0.56(12)	&	0.9(2)	&		&		&		\\
$^{196}$Pt	&		&	+0.604(48)	&	+0.54(9)	&	0.9(2)	&		&		&		\\
$^{198}$Pt	&		&	+0.63(2)	&	+0.61(11)	&	1.0(2)	&		&		&		\\
\hline\hline
\end{tabular}
\end{table}

In Table \ref{mu}, we list all the experimentally available $\mu(2^+_1)$ and $\mu(2^+_2)$ values as well as the corresponding TRM estimates. The magnetic-moment ratios of $\mu(2^+_2)/\mu(2^+_1)$ are also presented to demonstrate the accuracy of the  $\mu(2^+_1)=\mu(2^+_2)$ correlation predicted by the AHV. One sees that most nuclei tend to have experimental $\mu(2^+_2)/\mu(2^+_1)$ ratios close to 1 within experimental error, in general agreement with the AHV description, except for  $^{92}$Zr and $^{184,~186}$W. The $2^+_1$ and $4^+_1$ states of $^{92}$Zr have been assigned as non-collective $(\nu 1d_{5/2})^4$ configurations beyond the $N=50$ major shell \cite{mu-92zr}.
The abnormality of $\mu(2^+_2)$ for $^{184,~186}$W  has long been noted, with a hint at shape mixing, but still remains an open question \cite{ab-g-w-1,ab-g-w-2}.  Therefore, these nuclei should be neither AHV  nor TRM-applicable nuclei. It should also be noted that the TRM gives $\mu$ values in rough agreement with experiment, even though we are using somewhat oversimplified $g_R$ and $g_K$ estimates in Eq. (\ref{trm-mu-az}). The experimental $\mu(2^+_1)=\mu(2^+_2)$ correlation is not a natural result of the TRM except if $g_R=g_K$. Thus, one has to require a general $g_R\sim g_K$ relation to achieve a more realistic description in the TRM framework.

\subsection{E2 collectivity}\label{subsec-q}
We now focus on the correlation between E2 transition rates and the quadrupole moments of $2^+_1$ and $2^+_2$ states. According to Eq. (\ref{trm-q}), such a correlation in the TRM can be analytically expressed as:
\begin{equation}\label{trm-be2-q}
\begin{aligned}
&{\rm B(E2,~}2^+_1\rightarrow0^+_1)+{\rm B(E2,~}2^+_2\rightarrow0^+_1)\\
&=\frac{7}{10}~{\rm B(E2,~}2^+_2\rightarrow2^+_1)+\frac{49}{64\pi}~Q^2(2^+_1)\\
&\simeq 0.7~{\rm B(E2,~}2^+_2\rightarrow2^+_1)+0.244~Q^2(2^+_1).
\end{aligned}
\end{equation}
This equation is an alternative representation of the triangle relations proposed in Ref. \cite{trm-os}.

For the AHV, conside
ring $\langle 1||\tilde{b}||2\rangle^2=2\langle 0||\tilde{b}||1\rangle^2$ we can derive a similar formula to Eq. (\ref{ahv-q}), namely
\begin{equation}\label{ahv-be2-q}
\begin{aligned}
&{\rm B(E2,~}2^+_1\rightarrow0^+_1)+{\rm B(E2,~}2^+_2\rightarrow0^+_1)\\
&=\frac{5}{10}~{\rm B(E2,~}2^+_2\rightarrow2^+_1)+\frac{35}{64\pi}~Q^2(2^+_1)\\
&\simeq0.5~{\rm B(E2,~}2^+_2\rightarrow2^+_1)+0.174~Q^2(2^+_1)~.
\end{aligned}
\end{equation}

It is worth noting that  Eqs. (\ref{trm-be2-q}) and (\ref{ahv-be2-q})  both belong to the Kumar-Cline sum rules \cite{qsr-1,qsr-2}. The generalization of these sum rules may be expressed as
\begin{equation}\label{fit-be2-q}
\begin{aligned}
&{\rm B(E2,~}2^+_1\rightarrow0^+_1)+{\rm B(E2,~}2^+_2\rightarrow0^+_1)\\
&=c_1~{\rm B(E2,~}2^+_2\rightarrow2^+_1)+c_2~Q^2(2^+_1),
\end{aligned}
\end{equation}
where $c_1$ and $c_2$ are free variables. We perform a multiple linear fitting of Eq. (\ref{fit-be2-q}) to  all available experimental data from the ENSDF with $c_1$ and $c_2$ as fitting parameters. All told, 78 nuclei are considered in this fit, with the final correlation coefficient $R=0.980$, being very close to 1. This demonstrates that B(E2) values between ground states and low-lying $2^+$ stat
e
are highly correlated with $Q(2^+_1)$ in the ENSDF, as expected in the TRM and AHV. The best-fit results are $c_1=0.479$ and $c_2=0.188$. It seems that Eq. (\ref{ahv-be2-q}), {\it i.e.} the AHV expression, gives closer agreement with experiment.

\begin{figure}
\includegraphics[angle=0,width=0.45\textwidth]{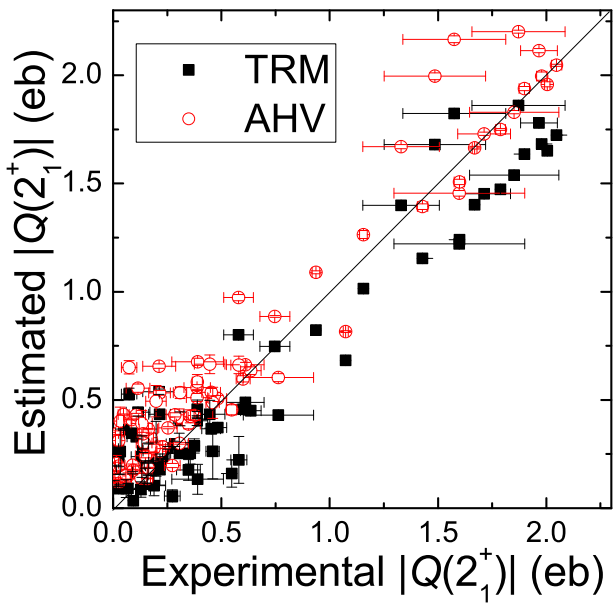}
\caption{(color online)  Plot of the estimated $|Q(2^+_1)|$ values based solely on the use of experimental B(E2) values according to Eqs. (\ref{trm-be2-q}) and (\ref{ahv-be2-q}), against the experimental data, for the 78 nuclei with available data
 in the ENSDF. The diagonal line is a measure of the quality of the estimates.}\label{be2-q}
\end{figure}

Based on Eqs. (\ref{trm-be2-q}) and (\ref{ahv-be2-q}), we estimate the magnitudes of $Q(2^+_1)$  using experimentally available B(E2) values in the TRM and AHV frameworks,  and then compare them with experiment. As for the linear fit of Eq. (\ref{fit-be2-q}), only 78 nuclei in the ENSDF enable such comparison. In Fig. \ref{be2-q}, we plot the $|Q(2^+_1)|$ values that emerge for these 78 nuclei  in comparison with the experimental values. The data points of both models scatter fairly closely around the diagonal line, supporting the validity of both the TRM and AHV approaches as global descriptions of low-lying E2 collectivity.  We also note that the TRM tends to give smaller $|Q(2^+_1)|$ values than experiment, whereas the AHV tends  to give larger values.

Spectroscopic properties of the low-lying states may also be  used to obtain alternative estimates for the quadrupole moments just discussed. In the TRM, excitation energies of the $2^+_1$, $2^+_2$ and $4^+_1$ states determine $\Gamma$, while the B(E2, $2^+_1\rightarrow 0^+_1$) and B(E2, $2^+_1\rightarrow 0^+_1$) determine $\gamma$ and $Q_0$  [see Eq. (\ref{trm-q})]. These five pieces of experimental information can be used to estimate the magnitude of $Q(2^+_1)$, using Eq. (\ref{trm-q}).  In the AHV framework, the wave-function amplitudes, $a_1$ and $a_2$ can be expressed in terms of the experimental excitation energies of the $2^+_1$ and $2^+_2$ states according to
\begin{equation}\label{a1a2}
a_1^2=\frac{2E(2^+_2)-E(2^+_1)}{3[E(2^+_2)-E(2^+_1)]},~a_2^2=\frac{E(2^+_2)-2E(2^+_1)}{3[E(2^+_2)-E(2^+_1)]}~.
\end{equation}
Using Eq. (\ref{ahv-q}), we can then estimate the magnitude of $Q(2^+_1)$ in the AHV as
\begin{widetext}
\begin{equation}\label{ahv-e-q}
|Q(2^+_1)|=\frac{16\sqrt{[2E(2^+_2)-E(2^+_1)][E(2^+_2)-2E(2^+_1)]}}{3[E(2^+_2)-E(2^+_1)]}\times\sqrt{\frac{\pi[{\rm B(E2,~}2^+_1\rightarrow 0^+_1)+{\rm B(E2,~}2^+_2\rightarrow 0^+_1)]}{35}}.
\end{equation}
\end{widetext}

\begin{figure}
\includegraphics[angle=0,width=0.45\textwidth]{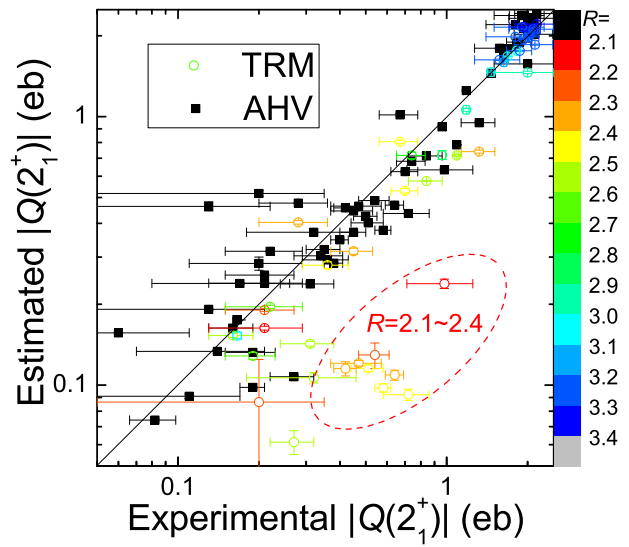}
\caption{(color online) Estimated $|Q(2^+_1)|$ values based on experimental excitation energies and B(E2)s (see text) against experimental values, for the 48 and 63 nuclei with available data in the ENSDF appropriate to the TRM and AHV, respectively. The diagonal line is a
 measure of the quality of the agreement. The TRM data points are shown in colors associated with the $R$ values (the map illustrated on the right). The red dashed ellipse, which highlights the TRM estimates that deviate most dramatically from the data. These deviations correspond to fairly typical vibrators with $R=2.1\sim 2.4$, including several Ru, Pd, and Cd isotopes}\label{e-q}
\end{figure}

There are 48 and 63 nuclei in the ENSDF for which a
 comparison between experimental $|Q(2^+_1)|$ values and their spectroscopically-based estimates are possible using the TRM and AHV, respectively. We plot these estimates against the experimental values in Fig. \ref{e-q}. We wish to emphasize here that the experimental ensembles of nuclei considered in Figs. \ref{be2-q} and \ref{e-q} are different.

In Fig. \ref{e-q}, the AHV estimate is invariably near the diagonal line, confirming the validity of Eq. (\ref{ahv-e-q}). The majority of the TRM estimates also agree with experiment. However, for the vibrational Ru, Pd and Cd isotopes with $R=2.1\sim2.4$, large deviations can be observed, as highlighted by the red dashed ellipse.  In contrast, the TRM estimates based solely on B(E2) values [see Fig. \ref{be2-q}] seemed to work well even for vibrational nuclei. We conclude, therefore, that the TRM may be more suitable to regulate electromagnetic properties involved in E2 collectivity than spectra. We believe that this may be attributable to two possible origins. On the one hand, the spectral description of the TRM involves more low-lying levels than does the AHV. Besides the $E(2^+_1)$ and $E(2^+_2)$ values required in the AHV, our TRM estimate  needs  additional input o
n the experimental $E(4^+_1)$ value. Experimental incompleteness of the low-lying level scheme, single-particle motion, shape coexistence or $\gamma$-instability could all interfere with the TRM spectral description. A second possible origin concerns the fact that, as mentioned above, descriptions of E2 electromagnetic properties in both the AHV and TRM approaches satisfy the Kumar-Cline sum rules, which may perhaps exist independent of the detailed spectral behavior. Therefore, even if the low-lying TRM spectral description is not very accurate, the relation between B(E2) values and $Q(2^+_1)$ from the TRM, {\it i.e}. Eq. (\ref{trm-be2-q}), could still be preserved by the Kumar-Cline sum rules. This situation actually has been reported in TRM calculations for the Os isotope chain \cite{trm-os}: even though the TRM always provided higher $K^{\pi}=4^+$ bands than experiment, E2 matrix elements for ground states and low-lying $2^+$ states are still described reasonably well by the TRM.

\section{Possible relation between the TRM and AHV}\label{cor}
In Sec. \ref{sys}, we saw that the TRM and the AHV both describe fairly robustly the relations between experimental excitation energies, magnetic moments and E2 collectivity, especially for the $2^+$ states. Furthermore, both describe $2^+$ states in terms of 2$\times$2 matrices.  If the $2^+_1$ and $2^+_2$ states of a nucleus can be spectrally described both by the TRM and the AHV, there exists an orthogonal matrix $U$ with transformation angle $\theta$ for this nucleus given by
\begin{equation}\label{u}
U=\begin{pmatrix}
\cos\theta&\sin\theta\\
-\sin\theta&\cos\theta
\end{pmatrix},
\end{equation}
such that
\begin{equation}
H^{2^+}_{\rm TRM}=UH^{2^+}_{\rm AHV}U^{\rm T}.
\end{equation}

This raises the question of whether there might be an intrinsic relation between the TRM and AHV for realistic nuclei, whereby both models describe those nuclei within roughly the same model space. If so, there woul
d exist a {\it unique} orthogonal transformation that relates the basis with good $K$ quantum number in the TRM and the phonon basis of the AHV for nuclei that are both TRM- and AHV-describable, namely
\begin{equation}
\begin{bmatrix}
|K=0\rangle\\
|K=2\rangle
\end{bmatrix}
=
U
\begin{bmatrix}
|1\rangle\\
|2\rangle
\end{bmatrix}.
\end{equation}

To clarify whether the $U$ matrix, {\it i.e.} the $\theta$ angle, is globally unique or robust for realistic nuclei, we calculate the $\theta$ distribution. The existence of a single dominant peak of this distribution would hint at the uniqueness or robustness of the orthogonal transformation between the TRM and AHV bases.  The calculation proceeds as follows. First, we extract the $\theta$ angles of the $U$ matrixes according to experimental excitation energies and the detailed structure of $H^{2+}_{\rm TRM}$ and $H^{2+}_{\rm AHV}$. In Sec. \ref{spe}, 203 nuclei in the ENSDF were determined to be TRM-solvable, and their $R$ distribution is plotted in Fig. \ref{trm-r}. We also find that these nuclei always have $E(2^+_2)>2E(2^+_1)$, and thus are AHV-describable. Therefore, these 203 nuclei define the largest ensemble of TRM \& AHV-describable nuclei in the ENSDF, and our $\theta$ distribution is calculated for this ensemble. Here, we note that for each nucleus in this ensemble there in fact exist two $\theta$ values, since the sign of the $\lambda$ value in $H^{2+}_{\rm AHV}$, {\it i.e.} Eq. (\ref{ahv-h}), can not be determined solely from excitation energies, so that a single level scheme from a given nucleus leads to two possible $\lambda$ values and thus two $\theta$ values. We also consider the $\theta$ magnitude only to simplify our analysis, since $\theta$ and $-\theta$ define the same transformation (if we change the phases of $|K=2\rangle$ and $|2\rangle$ in their respective bases). Second, we perform two-dimensional counting statistics for $\theta$ and $R=E(4^+_1)/E(2^+_1)$ for these 203 nuclei. This is because the $|\theta|$ and  $R$ values are both extracted from the same low-lying spectrum for a given nucleus. Thus, the $|\theta|$ value should most likely be correlated with $R$. In the ensemble of TRM-solvable nuclei, the $R$ distribution has its own bias as shown in Fig. \ref{ahv-e}, which may therefore also lead to dominant peaks in the $|\theta|$ distribution. By using the two-dimensional counting number, $N(|\theta|, R)$, we can decouple the potential correlation between $|\theta|$ and $R$ by defining for the  $|\theta|$ distribution
\begin{equation}\label{dis-q}
P(|\theta|)=\sum_R\frac{N(|\theta|,R)}{N(R)},
\end{equation}
where  $N(R)$ is the counting number of the $R$ distribution as shown in Fig. \ref{trm-r}. A dominant peak of such a calculated $P(|\theta|)$ distribution should avoid a false interpretation of the $|\theta|$ distribution due to any $R$ predominance, and thus signal a true robustness of the orthogonal transformation between the TRM and AHV basis.

\begin{figure}
\includegraphics[angle=0,width=0.45\textwidth]{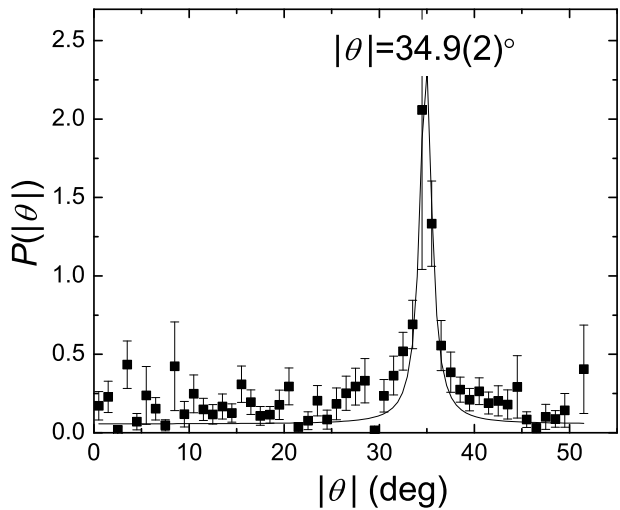}
\caption{ $R$-normalized $P(|\theta|)$ distribution (square points) defined by Eq. (\ref{dis-q}). The error bar represents the statistical error. The peak fit ( solid line) has its center at $|\theta|=34.9(2)^{\circ}$, as highlighted. This figure uses the same statistical ensemble as in Fig. \ref{trm-r}.}
\label{theta}
\end{figure}

The calculated $P(|\theta|)$ distribution  is plotted in Fig. \ref{theta}. There indeed exists a fairly narrow dominant peak  at $|\theta|=34.9(2)^{\circ}$, suggesting that most TRM and AHV Hamiltonian matrices are connected by a similar orthogonal transformation, and thus that their model spaces tend to be roughly the same for most nuclei.

This might provide a hint as to why these two simple, but seemingly very different, models are able to provide similar systematics for such a wide range of realistic nuclei, raising the question, therefore,  of whether the collective rotational characteristics exhibited by nuclei throughout much of the periodic table are indeed a reflection of an underlying rotor behavior or perhaps have a deeper origin.

In this spirit, it is interesting to note that we have found a somewhat analogous  behavior in the IBM-1 \cite{ibm} (the IBM with one type of boson) when there are only two bosons present. Governed by a U(6) symmetry, the IBM-1 naturally incorporates several types of collectivity through its various dynamical symmetry limits.  For a perfect harmonic vibrator, the U(5) limit applies, whereby one-phonon and two-phonon states with spin 2$\hbar$ involve one and two $d$ bosons, respectively. Mixing of these states, {\it i.e.} phonon-state mixing, gives the AHV model described herein. If the mixing is driven by an SU(3) Hamiltonian, we find that a {\it unique} linear transformation of such phonon states provides two rotational $2^+$ states, as defined in Eq. (\ref{u}). By using an IBM-1 code, we have numerically calculated the transformation angle between the states of this basis and the lowest $2^+$ states of the SU(3) basis, and find a rotation angle of $\theta= 28.1^{\circ}$. This is reasonably close to the peak of $\theta$ distribution f
or realistic nuclei as shown in Fig. \ref{theta}. It should be noted, however, that the TRM does not derive from the SU(3) limit of the IBM-1, which is an axially symmetric rotor,  and thus a direct connection of the two angles that emerge cannot be made. Nevertheless, the results are sufficiently intriguing to suggest the need for further work to explore this issue.

\section{summary}\label{sum}
To summarize, we have studied two simple models of nuclear structure, the TRM and the AHV, both of which have the feature that they naturally satisfy the property $Q(2^+_1)= -Q(2^+_2)$ that has been found globally for nuclei exhibiting quadrupole collectivity. We have shown that both of these models can describe systematic correlations among excitation energies, magnetic moments and E2 collectivity, especially for the lowest two $2^+$ states, across a wide range of the periodic table, which furthermore are in general agreement with experimental data. For the few exceptions where agreement is not achieved, we provide plausible explanations. The correlations provided by these models could prove useful as a way to predict data where experiment is not available or to ``verify" existing data. We also find that the TRM and AHV Hamiltonian matrixes can be connected by an orthogonal transformation which seems to be roughly the same for
 most nuclei. This seems to hint that the TRM and AHV, though seemingly quite different models of nuclear collective behavior - one based on a rotor and the other not -  may in fact share the same model space for realistic nuclear systems, a conjecture that we believe deserves further theoretical investigation.

\acknowledgements
One of the authors (YL) wishes to thank an anonymous referee of Ref. \cite{lei} for reminding him of the AHV explanation for the $Q(2^+_1)=-Q(2^+_2)$ correlation. Extensive discussions with Dr. J. M. Allmond and Prof. N. Yoshida are also gratefully acknowledged. This work was supported by the National Natural Science Foundation of China under Grant No. 11305151 and the Research Fund for the Doctoral Program of the Southwest  University of Science and Technology under Grant No. 14zx7102.

\end{document}